\documentclass[10pt, a4paper,final]{amsart}
\usepackage{amssymb}
\usepackage{amsthm}
\usepackage{graphics}
\usepackage[dvips]{graphicx}
\newtheorem{thm}{Theorem}[section]

\newtheorem{lem}[thm]{Lemma}

\theoremstyle{definition}

\theoremstyle{remark}

\numberwithin{equation}{section}

\newcommand{\ket}[1]{\mbox{$|#1\rangle$}}

\newcommand{\inner}[2]{\mbox{$\langle#1|#2\rangle$}}

\newlength{\defbaselineskip}
\newcommand{\setlinespacing}[1]%
           {\setlength{\baselineskip}{#1 \defbaselineskip}}

\begin{document}

\title{A Tight Bound for Probability of Error for Quantum Counting Based Multiuser Detection}%
\author{S\'andor Imre, Ferenc Bal\'azs}%
\address{Budapest University of Technology and Economics\\
Department of Telecommunications\\ Mobile Communications
Laboratory \\ H-1117 Budapest, Magyar Tud\'osok krt. 2., HUNGARY}%
\email{IMRE@hit.bme.hu - BALAZSF@hit.hit.bme.hu}%

\keywords{Multiuser detection, Quantum Counting, Grover's Algorithm, Quantum computing, Quantum Signal Processing}%

\date{}%

\maketitle
\begin{abstract} Future wired and wireless
communication systems will employ pure or combined Code Division
Multiple Access (CDMA) technique, such as in the European 3G
mobile UMTS or Power Line Telecommunication system, but also
several 4G proposal includes e.g. multi carrier (MC) CDMA. Former
examinations carried out the drawbacks of single user detectors
(SUD), which are widely employed in narrowband IS-95 CDMA systems,
and forced to develop suitable multiuser detection schemes to
increase the efficiency against interference. However, at this
moment there are only suboptimal solutions available because of
the rather high complexity of optimal detectors. One of the
possible receiver technologies can be the quantum assisted
computing devices which allows high level parallelism in
computation. The first commercial devices are estimated for the
next years, which meets the advert of 3G and 4G systems. In this
paper we analyze the error probability and give tight bounds in a
static and dynamically changing environment for a novel quantum
computation based Quantum Multiuser detection (QMUD) algorithm,
employing quantum counting algorithm, which provides optimal
solution.
\end{abstract}

\section{Introduction}
The subscribers of next generation wireless systems will
communicate simultaneously, sharing the same frequency band. All
around the world 3G mobile systems apply Direct Sequence - Code
Division Multiple Access (DS-CDMA) promising high capacity and
inherent resistance to interference, hence it comes into the
limelight in many communication systems. Nevertheless due to the
frequency selective property of the channel, in case of CDMA
communication the orthogonality between user codes at the receiver
is lost, which leads to performance degradation. Single-User
detectors were overtaxed and shown rather poor performance even in
multi-path environment \cite{verdu}. To overcome this problem,
recent years multiuser Detection (MUD) has received considerable
attention and become one of the most important signal processing
task in wireless communication. \par Verdu \cite{verdu} has proven
that the optimal solution is consistent with the optimization of a
quadratic function, which yields in MLSE (Maximum-Likelihood
Sequence Estimation) receiver. However, to find the optimum is an
\textit{NP}-hard problem as the number of users grows. Many
authors proposed sub-optimal linear and nonlinear solutions such
as Decorrelating Detector, MMSE (Minimum Mean Square Error)
detector, Multistage Detector, Hoppfield neural network or
Stochastic Hoppfield neural network \cite{verdu,dejou00,var90},
and the references therein. One can find a comparison of the
performance of the above mentioned algorithms in \cite{dejou01}.
\par Nonlinear sub-optimal solutions provide quite good
performance, however, only asymptotically. Quantum computation
based algorithms seem to be able to fill this long-felt gap.
Beside the classical description, which we recently use,
researchers in the early $20^{th}$ century raised the idea of
quantum theory, which nowadays becomes remarkable in coding
theory, information theory and for signal processing
\cite{pres98}. \par Nowadays, every scientist applies classical
computation, using sequential computers. Taking into account that
Moore's law can not be held for the next ten years because silicon
chip transistors reach atomic scale, therefore new technology is
required. Intel, IBM, AT\&T and other companies invest large
amount of research to develop devices based on quantum principle.
Successful experiments share up that within 3-4 years quantum
computation (QC) assisted devices will be available on the market
as enabling technology for 3G and 4G systems \cite{imr01b,imr01c}.
\par This paper is organized as follows: in Section
\ref{sec:back} we shortly review the necessity of multiuser
detection, as well as the applied quantum computation method is
shown. In Section \ref{sec:model} the proposed quantum multiuser
detector model is introduced. Furthermore, in Section
\ref{sec:error}. we give and proof a tighter probability error for
our model. Finally we conclude our paper in Section
\ref{sec:conc}.

\section{Theoretical Backgrounds}\label{sec:back}
\subsection{Multiuser Detection}\label{sec:mud} One of the major
attributes of CDMA systems is the multiple usage of common
frequency band and the same time slot. Despite the interference
caused by the multiple access property, the users can be
distinguished by their codes. Let us investigate an uplink
DS-DCDMA system, where the $i^{th}$ symbol of the $k^{th}$
$(k=1,2,\dots,K)$ user is denoted by $b_k(i)$, $b_k(i)\in \{+1,
-1\}$. In DS-CDMA systems an information bearing bit is encoded by
means of a user specific code with length of the processing gain
(\textit{PG})\cite{verdu}. In case of uplink communication we
assume perfect power control. In the receiver side it is not
required synchronization between input signals and user specific
codes, however we make our decision on symbols. Applying BPSK
modulation, the output signal of the $k^{th}$ user, denoted by
$q_k(t)$, is given as
\[q_k(t)=\sqrt{E_k}b_k(i)s_k(t),\]
where $E_k$ and $s_k(t)$ are the energy associated to the $k^{th}$
and the user continuous signature waveform, respectively
\[s_k(t)=\sum_{j=0}^{PG-1}c_k(j)g_k(t-jT_c),\] $T_c$ denotes
the time duration of one chip, $c_k(j)$ is the $j^{th}$ chip of
the code word of subscriber $k$ and $g_k(t)$ refers to the chip
elementary waveform. We investigate a one path uplink wideband
CDMA propagation channel. The channel distortion for the $k^{th}$
user is modelled by a simple impulse response function
$h_k(t)=a_k\delta(t-\tau_{k})$, where $a_{k}$ and $\tau_{k}$ are
the path gain and the delay of the $k^{th}$ user, respectively
\cite{liu}. They are assumed to be constant during a symbol period
of $T_s$. This model contains almost all elements of a typical
WCDMA channel except multipath propagation, which was omitted to
simplify the explanation of the new quantum computation based
multiuser detection scheme. However, based on the results of the
present paper, multipath propagation can be included into the
channel model on a very simple way.
\par The received signal is the sum of arriving signals plus a
Gaussian noise component and thus can be written as follows:
\begin{eqnarray}
r(t)=\sum_{k=1}^Kh_k(t)\ast q_k(t)+n(t) \nonumber
=\sum_{k=1}^K\sqrt{E_k}a_{k}b_ks_k(t-\tau_k)+n(t), \label{eq:mud1}
\end{eqnarray}
where $K$ is the number of users using the same band, $n(t)$ is a
white Gaussian noise with constant $N_0$ spectral density.
\par Unfortunately, the search for the globally optimal MUD
\cite{verdu} usually proves to be rather tiresome, which prevents
real time detection (its complexity by exhaustive search is
$\mathcal{O}$$(2^K)$). Therefore, our objective is to develop new,
powerful detection technique, which paves the way toward real time
MUD even in highly loaded system. Since classical multiuser
detection schemes only try to minimize the probability of error in
noisy and high interference environment, they, even also optimal
solutions, can commit an error. Actually, these classical
approaches make compromize between computational complexity,
probability of error and time barrier required for efficient
working. On the other hand, QMUD provides for typical CDMA systems
$\mathrm{BER}$$\approx 10^{-10}$ and it is able to indicate
undistinguishable decision situations for correction by higher
layer protocols \cite{imr01}.

\subsection{Quantum Computation Theory}\label{sec:qt}
\par Quantum theory is a mathematical model of a physical system. To
describe such a model one needs to specify the representation of
the system. According to the axioms of quantum mechanics, every
physical system can be characterized by means of its states
$\ket{\varphi}$\footnote{Say ket $\varphi$ (using Dirac's
notation).} in the \textit{Hilbert} vector space over the complex
numbers $\mathbb{C}$, whereas a physical quantity can be described
as Hermitian operator $U=U^\dag$, respectively. 
\par In the classical information theory the smallest  information conveying unit
 is the \textit{bit}. The counterpart unit
in quantum information is called the \textit{"quantum bit"}, the
qubit. Its state can be described by means of the state
$\ket{\varphi}$, $\ket{\varphi}=\alpha\ket{0}+\beta\ket{1}$, where
$\alpha,\beta \in \mathbb{C}$ refers to the complex probability
amplitudes, where $|\alpha|^2+|\beta|^2=1$ \cite{shor98,pres98}.
The expression $|\alpha|^2$ denotes the probability that after
measuring the qubit it can be found in computational base
$\ket{0}$, and $|\beta|^2$ shows the probability to be in
computational base $\ket{1}$. In more general description an
$N$-bit \textit{"quantum register"} (qregister) $\ket{\varphi}$ is
set up from qubits spanned by $\ket{i}$ $i=0\dots(N-1)$
computational bases, where $N=2^n$ states can be stored in the
qregisters at the same time \cite{imr01}, describing
\begin{eqnarray}
\ket{\varphi}=\sum_{i=0}^{N-1}\varphi_i\ket{i} & \varphi_i\in
\mathbb{C}, \label{eq:6}\end{eqnarray} where $N$ denotes the
number of states and $\forall i\neq j$, $\inner{i}{j}=0$,
$\inner{i}{i}=1$, $\sum|\varphi_i|^2=1$, respectively. It is worth
mentioning, that a  transformation $U$ on a qregister is executed
parallel on all $N$ stored states, which is called \textit{quantum
parallelization}.

\section{Quantum multiuser Detector Employing Quantum Search}\label{sec:model}
For the optimal decision it would be necessary a fully
comprehensive knowledge about the symbols sent by all the
subscribers in the coverage area of base station, a realization of
the delay and noise, which is typical for a particular
communication channel and all the user specific codes. All this
information cannot be stored in a single database, which should be
build just one time at all. In pursuance of detection one compares
the quantized received signal with the content of the database.
This task can be done efficiently, employing the Grover
\cite{gro96, gro00} database search algorithm $(\mathcal{G})$,
which is proved to be optimal \cite{zal99} in
$\mathcal{O}(\sqrt{N})$ steps, where $N$ denotes the length of the
database. \par The optimal number of iterations is depending on
the initial angle $\theta$ between the initial state
$\ket{\gamma}$ and $\ket{\alpha}$, as well as the number of the
identical entries $M$ in the database. In \cite{gro96} it was
shown that the rotation of state $\ket{\gamma}$ to the desired
state $\ket{\beta}$ after $d$ evaluations of Grover operator
$(\mathcal{G})$ is
\begin{equation}
\mathcal{G}^{\mathit{d}}\ket{\gamma}=\cos\left(\frac{\mathrm{2}\mathit{d}+\mathrm{1}}{\mathrm{2}}\theta\right)\ket{\alpha}+\sin\left(\frac{\mathrm{2}\mathit{d}+\mathrm{1}}{\mathrm{2}}\theta\right)\ket{\beta},
\end{equation}
whereas in \cite{zal99} it was proved that the optimal number of
iterations $l$ is given as
\begin{equation}
d=\mathrm{floor}\left.|\left(\frac{\arccos\sqrt{\frac{\mathit{M}}{N}}}{\theta}\right)\right|_{d=x.5\rightarrow
d=x}.
\end{equation}
In case of $M\ll N$ or even it exists no multiple entries in the
qregister the angle
\begin{equation}
\theta\simeq\sin\theta=2\sqrt{\frac{M}{N}},
\label{eq:3.3}
\end{equation}
where the optimal number of evaluations is equal to
$d\simeq\frac{\pi}{4}\sqrt{\frac{M}{N}}$ \cite{zal99}. It is worth
mentioning that in case of $N=4M$, the initial angle
$\theta=60^{\circ}$, which leads to only one evaluation. \par In
case of quantum detection, however, it is not required to know the
exact position of the representative of the received signal in the
qregister, only whether is it contained in the database or not.
For this purpose, the so called \textit{Quantum Counting} based on
Quantum Fourier Transformation (QFT) designed for phase estimation
is well suited. In the following part of this section we introduce
the buildup of the applied databases, qregisters. Furthermore, we
shortly discuss the quantum counting problem and our way to deal
with quantum multiuser detection problem.
\subsection{Applied Databases and Qregisters}
Let us assume we would like to detect the $k^{th}$ user's symbol,
denoted by $b_k$. To design a suitable multiuser detector all of
the user specific code words $\mathbf{c}_k$ have to be known at
the base station, collected in a $(PG \times K)$ sized database
$\mathcal{C}=[\mathbf{c}_1, \mathbf{c}_2, \ldots,
\mathbf{c}_{\textit{K}}]$. In addition we create a qregister
$\ket{\varphi}$, which contains all the possible received signal
configurations $\ket{x}$ without $b_k$ \begin{equation}
\ket{\varphi}=\sum_{i=0}^{N-1}\varphi_i(x)\ket{x}, \label{eq:x}
\end{equation} where $\ket{x}$ are the computational base states.
Since every possible noise and delay states for a given
transmitted bit is involved in the prepared register, it is
superfluous to send any phase information, which allows
non-coherent detection. The first $(K-1)$ qubits represent all the
combination of the symbols transmitted by all the interferer, e.g.
if the value of $(K-j)^{th}$ qubit in a given $\ket{x}$ is equal
to 1, it shows that the user $(K-j)$ has been transmitted a symbol
denoted by $b_{K-j}=1$, otherwise if $b_k=-1$ the qregister has an
entry $0$. For a proper utilization of the qregister the next two
parts in $\ket{x}$ is reserved for all quantized state of the
noise $n(t)$ and the delay $\tau$: $0 \leq \tau \leq T_s$. For the
sake of simplicity we assume to apply a linear quantization
$n=i\cdot n_R$, where $i\in [-N_n, +N_n]$ and $n_R$ refers to the
quantization step. It is remarkable, if the \textit{probability
density function} (pdf) of the target quantity is known, such as
by the Gaussian noise, an adequate nonlinear quantization could be
used, whereby most significant values can be sampled denser, which
can decrease the number of required qbits for description of
noise. The values of the delay $\tau$ can be generated in a
straitforward manner, $\tau=j\cdot\tau_R$, where $j$ could be
$j\in \left[0,\frac{T_s}{\tau_R}\right]$. With the above mentioned
description the size of qregister $\ket{\varphi}$ is given by
\begin{equation}N_{qreg}=\lceil\mathrm{ld} \mathit{N}\rceil=(K-1)+\left\lceil\mathrm{ld}
(2\mathit{N_n}+1)+\mathrm{ld}\left(\frac{\mathit{T_s}}{\tau_{\mathit{R}}}+1\right)\right\rceil,
\label{eq:21}
\end{equation}
where $\lceil\cdot\rceil$ denotes the round up operation to the
next integer. We consider two independent qregisters
$\ket{\varphi_{+1}^k}$ and $\ket{\varphi_{-1}^k}$ for every user
$k$ according to sent information symbol $+1$ or $-1$. From this
point we just refer to $\ket{\varphi_{j}^k}$, where $j=\{-1,1\}$,
however, the operations on each qregister can be done in the same
manner. Thus, $\ket{\varphi_j^k}$ are the databases as well as
detection means to check whether the received signal can be found
in the databases \cite{imr01c}.
\subsection{Quantum
Counting}\label{ssec:counting} 
\begin{figure}[tb]
\begin{center}
\includegraphics[width=90mm, height=30mm]{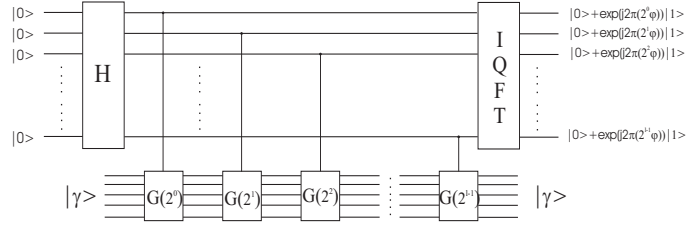}
\end{center}
\caption{The quantum counter circuit}\label{fig:counter}
\end{figure}

Performing multiuser detection it
is obvious not to look for a pattern in a database itself but to
indicate whether or not a solution even exists, consequently, the
number of solutions have to be summed up. This quantum counting
algorithm can be accomplished with combining the Grover quantum
search algorithm mentioned in Section \ref{sec:model}, with the
phase estimation based on the quantum Fourier transformation
\cite{nielsen}. The quantum counting algorithm is able to estimate
the eigenvalues of a unitary operation $U$, or in this case the
Grover iterations $(\mathcal{G})$
\begin{eqnarray}
\mathcal{G}\ket{\mathit{\gamma}}=\mathrm{e}^{\mathit{j}\theta}\ket{\gamma},
& \theta=2\pi\delta,\label{eq:3.6}
\end{eqnarray} which is in a close connection with the demanded
number of solutions $M$ on the search. In addition to the original
Grover circuit a new register $\ket{t}$ of $l$ new qubits should
be feed to the quantum counting algorithm, whose state is
initialized to an equal superposition of all possible inputs by a
Hadamard gate, as in Figure \ref{fig:counter}. The vector
$\ket{\gamma}$, which contains all the possible signal states with
equally distributed probability amplitudes, is build up in the
same way. The original Grover block $\mathcal{G}$ becomes a
controlled Grover block, where the transformation (a rotation with
angle $\theta$, such as in the original Grover algorithm) will be
only evaluated if the control signal equals to one. The goal of
quantum counting is to estimate $\delta$ to $m$ bits of accuracy,
with probability of success $1-\varepsilon$. The state of the
register $\ket{t}$ after the controlled Grover iterations is
\begin{equation}
\frac{1}{2^{l/2}}\sum_{k=0}^{2^{l-1}}e^{j2\pi
k\delta}\ket{k}\ket{\gamma},\label{eq:23}
\end{equation}
which is exactly the QFT belonging to $\theta$. Hence, to estimate
$\delta$ an inverse quantum Fourier transformation is executed on
(\ref{eq:23}) before the measurement. In case of $\delta
>0$ after a measurement, the demanded pattern is contained in the qregister
$\ket{\varphi_j^k}$, and if it is equal to zero, it is not
included in. \section{The required number of additional
qubits}\label{sec:error}\par The required number of qubits $l$
\begin{equation}l=m+\left\lceil
\mathrm{ld}\left(2+\frac{1}{2\mathit{\varepsilon}}\right)+\mathrm{ld}\mathit{\pi}\right\rceil.\label{eq:4.1}\end{equation}
in $\ket{t}$, and hereby the required number of Grover blocks
$\mathcal{G}$ can be calculated in function of desired accuracy
$2^{-m}$ of $\delta$ and the probability $\varepsilon$ of false
detection. This error $\varepsilon$ is occurred if $\delta$ is not
in form of $z/2^l$, where $z\in0,1,\ldots, 2^l$. Hence it can be
regarded as the probability of false detection. \lem The accuracy
$m$ of $\delta$ can be given as
\begin{equation} \sup
\{m\}=\left\lceil\frac{N_{qreg}}{2}-1\right\rceil.
\label{eq:4.2}\end{equation} \proof Fortunately, we are able the
upperbound $m$ in (\ref{eq:4.1}). From the accuracy point of view
the worst case is occurred if $M=1$, because the initial angle
$\theta$, yielding form (\ref{eq:3.3}), is the smallest. Hence,
$m$ should be chosen as large as possible, which leads to
\[\theta_{\min}=
2\sqrt{\frac{1}{N}}=2\sqrt{\frac{1}{2^{N_{qreg}}}}=2^{\left(-\frac{N_{qreg}}{2}+1\right)}\Rightarrow
\sup \{m\}=\left\lceil\frac{N_{qreg}}{2}-1\right\rceil
\] \qed
\par Therefore, we have a direct connection between the detection
error probability and the required length of register $\ket{t}$.
However, before the measurement an inverse discrete QFT is done,
which adds additional states beside $\ket{\varphi}$ with a given
probability, if $\theta$ is not an integer power of two, that
could lead to false detection in the receiver. Hence, the
expression
$p=\left\lceil\mathrm{ld}\left(2+\frac{1}{2\mathit{\varepsilon}}\right)\right\rceil$
\cite{nielsen} in (\ref{eq:4.2}) is just a approximation. In
general, $p$ is needed to represent $\theta$ more precisely, and
thus to decrease the probability of false detection, but the
accuracy of the estimated angle $\widetilde{\theta}$ is $2^{-m}$
in the future too. This means, that $p$ only influences the
probability amplitudes, which leads to be enough measuring the
first $m$ most significant bits!
\subsection{Methods to improve the measurement}
\par To improve the proper error probability of detection, there are
some possibilities.
\begin{enumerate}
\item One should chose $\theta=2^L$, where $L\in \{\mathbb{N}\}$.
Unfortunately, except the trivial case, where $\theta=0$, which
means the desired state $\ket{r^k}$ is not in the qregister, it
happens very rarely.
\item Our goal is to distinguish the case $\theta=0$ from $\theta\neq
0$. It is known from the previous point, that the detection of
$\theta=0$ is certain. Consequently, if $\theta\neq 0$ a false
detection is occurred if and only if we decide for $\ket{0}$ after
the measurement. For a more precise estimation of the error
probability $P_{error}$, one should extend the size of the
qregister instead of $p$ bits, with $C$ more bits. We will show
that based on our results $C$ can be much smaller than $p$ at a
given $P_{error}$, where
\end{enumerate} \begin{equation}
P_{error}=P(\widetilde{\theta}=0|\theta\neq 0)\cdot P(\theta\neq
0)+P(\widetilde{\theta}\neq 0|\theta= 0)\cdot P(\theta= 0),
\label{eq:4.3}
\end{equation}
where $\widetilde{\theta}$ refers to the estimated phase. The
second part of the right side in (\ref{eq:4.3}) is equal zero,
because of the previous analysis. Assuming an unknown
\textit{a-priori} probability (\ref{eq:4.3}) becomes simpler
\[P_{error}\leq P(\widetilde{\theta}=0|\theta\neq 0).\]

The state of the examined $\ket{t}$ qregister before measurement
is given as
\begin{eqnarray}
\label{eq:4.4}
\frac{1}{\sqrt{2^l}}\sum_{k=0}^{2^{l}-1}\left[e^{j2\pi\delta
k}\underbrace{\frac{1}{\sqrt{2^l}}\sum_{i=0}^{2^{l}-1}e^{-j\frac{2\pi
ki}{2^{l}}}}_{\mbox{\scriptsize{IQFT corresponding to
$\ket{k}$}}}\ket{i}\right]=\frac{1}{\sqrt{2^l}}\sum_k\sum_ie^{j2\pi\delta
k}e^{-j2\pi ki}\ket{i}= \nonumber
\\=\frac{1}{\sqrt{2^l}}\sum_k\sum_ie^{j2\pi
k\left(\delta-\frac{i}{2^l}\right)}\ket{i}=\frac{1}{\sqrt{2^l}}\sum_i\sum_ke^{j2\pi
k\left(\delta-\frac{i}{2^l}\right)}\ket{i},
\end{eqnarray}
where $0\leq\delta\leq 1$ is a real number. Furthermore, the
probability amplitude of the state $\ket{i}$ is
\[
a_i(l)=\frac{1}{\sqrt{2^l}}\sum_{k=0}^{2^l-1}e^{j2\pi
k\left(\delta-\frac{i}{2^l}\right)}=\frac{1}{\sqrt{2^l}}\sum_k
\left(e^{j2\pi\left(\delta-\frac{i}{2^l}\right)}\right)^k,
\]
which is a geometrical series \cite{nielsen}
\begin{equation}
a_i(l)=\frac{1}{2^l}\left(\frac{e^{j2\pi\left(\delta
2^l-1\right)}-1}{e^{j2\pi\left(\delta
-\frac{i}{2^{l}}\right)}-1}\right). \label{eq:4.5}
\end{equation}
The demanded probability of error can be described as the sum of
over all probability amplitudes given in (\ref{eq:4.5})
\begin{equation}
P_{error}\leq P(\widetilde{\theta}=0|\theta\neq
0)=\sum_{i=0}^{2^C-1}\left|a_i(l)\right|^2,
\label{eq:4.6}\end{equation} form which the parameter $C$ is
calculable at a given $P_{error}$.
\subsection{Upper bound of error probability}\par While $C$ can not be expressed explicitly from (\ref{eq:4.6}), hence, in case of
 adaptive receiver structure it would be very useful to give less complex way of determine $C$. For this purpose an upper
bound has to be given for (\ref{eq:4.5}). It is considerable that
with increasing the size of the qregister with $C$ new bits, the
sum in (\ref{eq:4.6}) may also greater, however, due to more
accurate representation of $\theta$ the probability amplitude
belonging to the false vectors diminishes better and better.
\subsubsection{Numerator of (\ref{eq:4.5})}
For the numerator of (\ref{eq:4.5}) in \cite{nielsen} was shown an
unambiguous upper bound $\left|e^{j2\pi\left(\delta
2^{l}-1\right)}\right|$ $\leq 2$, which is simple the diagonal of
unit circle $\left|e^{j\alpha}\right|$. We show now, that under
certain conditions a tighter bound can be given. \thm \label{thm1}
For a fixed $\delta$ the numerator of the sum of the probability
amplitudes is
\[\max\left\{\pi\left(2^{m+c+1}\delta-\mathrm{floor}(2^{m+c+1}\delta)\right),
\pi-\pi\left(2^{m+c+1}\delta-\mathrm{floor}(2^{m+c+1})\right)\right\}.\]
\proof Unfortunately, the numerator of (\ref{eq:4.5}) is not
monoton on $\delta$ and $l$, hence, the maximum has to be searched
which is less than $2$. The unit circle is divided into two
regions, where $0\leq |\alpha| \mathrm{mod} 2\pi\leq \pi$ and
$\pi\leq |\alpha| \mathrm{mod} 2\pi\leq 2\pi$, where we should
find
\[\max_{l,\delta}\{|\alpha|\mathrm{mod} \pi\}=\max_{l,\delta}\{\alpha(l,\delta)\},\]
\[\max_{l,\delta}\{\pi-|\alpha|\mathrm{mod} \pi\}=\max_{l,\delta}\{\pi-\alpha(l,\delta)\},\] respectively,
whereby we mapped $|\alpha|$ in the range $[0,\pi]$, where
$|\alpha|$ strictly monoton. To get a proper upper bound the
maximum of the two regions for a fixed $\delta$
\begin{equation}\left[\left(2\pi
\left|2^t\delta-l\right|\right)\mathrm{mod}\right]\pi=\left[\left(2\left|2^t\delta-l\right|\right)\mathrm{
mod } 1\right]\pi.\label{eq:4.7}\end{equation}  The case of, when
$2^t\delta-l\leq 0$ in (\ref{eq:4.7}), yields in $l\geq
2^t\delta$, which can not come true, inasmuch as we sum up to
$2^C-1$, which is always less than $2^t$. Hence, the absolute
value operator can be neglected. The opposite case, where
$2^t\delta-l\geq 0$ in (\ref{eq:4.7}) the non-integer part becomes
$\pi\left(2^{t+1}\delta\right)\mathrm{ mod }1$ because of $2l$ is
an integer. These two considerations lead to \begin{equation}
\max\left\{\pi\left(2^{m+c+1}\delta-\mathrm{floor}(2^{m+c+1}\delta)\right),
\pi-\pi\left(2^{m+c+1}\delta-\mathrm{floor}(2^{m+c+1})\right)\right\}.\end{equation}\qed
\subsubsection{Denominator of (\ref{eq:4.5})}
Also in \cite{nielsen} was shown, that the denominator of
(\ref{eq:4.5}) $\left|e^{j2\pi\left(\delta
-\frac{i}{2^{l}}\right)}\right|=\left|e^{j\beta}\right|$ is always
greater than $\frac{2\left|\beta\right|}{\pi}$ in the range
$[-\pi,\pi]$. \thm \label{thm2} Employing the Grover search
algorithm in quantum counting the denominator of (\ref{eq:4.5})
becomes $\frac{2\sqrt{2}\left|\beta\right|}{\pi}$. \proof For the
proof, we should first evaluate $\beta_{\max}$, which is
\[\beta_{\max}=\min_l\max_{\delta}
2\pi\left(\delta_{\max}-\frac{l_{\min}}{2^t}\right).\] Form
(\ref{eq:3.6}), where the maximal value of $\theta=\frac{\pi}{2}$,
resulting from employing the Grover searching in the quantum
counting algorithm, $\delta_{\max}=\frac{1}{4}$. Furthermore, it
is obvious, that $l_{\min}=0$, which two considerations lead to
\[2\pi\left(\frac{1}{4}-0\right)=\frac{\pi}{2}.\] For the
complete examination, one should also look for the $\beta_{\min}$
value, which yields in
$2\pi\left(\delta_{\min}-\frac{2^C-1}{2^t}\right)$, where
$\delta_{\min}$ is very close to zero and the second part greater
than zero except for $C>t-2$, which is not realized almost surely.
Following the above mentioned ideas, $|\beta|\leq \frac{\pi}{2}$.
Applying $\beta_{\max}$ the denominator of (\ref{eq:4.5}) becomes
\begin{equation}
e^{j2\pi\left(\delta
-\frac{i}{2^{l}}\right)}-1=\frac{2\sqrt{2}\overbrace{\left|2\pi\left(\delta
-\frac{i}{2^{l}}\right)\right|}^{|\beta|}}{\pi}=4\sqrt{2}\left(\delta-\frac{l}{2^t}\right).
\end{equation}
\qed
\subsection{Simulation of new bounds}
\begin{figure}[tb]
\begin{center}
\includegraphics[width=90mm, height=60mm]{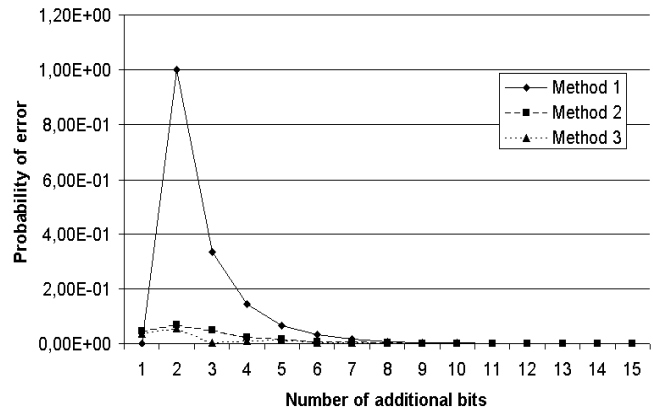}
\end{center}
\caption{The probability of error vs. additional
bits}\label{fig:2}
\end{figure}

To verify our results we performed a computer simulation of our
new probability of error upper bounds and compared them with them
of the model given in \cite{nielsen}, which is denoted by
\textit{Model 1} in Fig. \ref{fig:2}. For a static environment we
propose \textit{Method 3}, where an exact calculation of
(\ref{eq:4.6}) is done, whereas \textit{Method 2} refers to our
upper bound for dynamically changing system, following
\textit{Lemma 1}, \textit{Theorem 4.2} and \textit{Theorem 4.3}.
In Fig. \ref{fig:2} the probability of errors of the outcomes of
the three methods are depicted against the number of additional
bits $C$. It is noticeable a remarkable difference  between
\textit{Model 1} and \textit{Model 2 \& 3} especially at low
values of $C$. Conversely, the likewise less computational
complexity loaded \textit{Model 3} shows a quiet close result to
the theoretical one \textit{Model 2}. It is also observable that
the error vanish with very few new bits in the qregister.
\section{Conclusions}\label{sec:conc}
\par In this paper we presented a quantum computation based
multiuser detection algorithm, which involves a modified quantum
counting algorithm, employing the Grover's quantum search method.
We showed an exact computation method for the error probability,
which can be obtained in a static environment. Furthermore, we
also gave an fast computable approximation of the upper bound of
error probability very close to the theoretical one, which is
usable in adaptive receiver. The new method utilizes one of the
possible future receiver technologies of 3G and 4G mobile systems,
the so called quantum assisted computing. QMUD provides optimal
detection in finite time and complexity when classical methods can
achieve only suboptimal solutions. The proposed algorithm has
strong resistance against MAI and noise.

\end{document}